\documentclass[preprint,amsmath,amssymb,aps,prd,floatfix,]{revtex4-1}

\usepackage{graphicx}
\usepackage[normalem]{ulem}
\usepackage{dcolumn}
\usepackage{bm}
\usepackage{hyperref}
\usepackage{color}

\definecolor{ForestGreen}{RGB}{34,139,34}

\begin{document}
	
	\title{Cubic Galileon Gravity in the CMB}
	
	\author{Gen Ye}
	\email{ye@lorentz.leidenuniv.nl}
	\affiliation{Institute Lorentz, Leiden University, PO Box 9506, Leiden 2300 RA, The Netherlands	}
	\author{Alessandra Silvestri}%
	\affiliation{Institute Lorentz, Leiden University, PO Box 9506, Leiden 2300 RA, The Netherlands	}
	
	\begin{abstract}
		Among the models addressing the Hubble tension, those introducing a dynamical dark component around recombination have been the most promising thus far. Their study has highlighted that, in fact, cosmic microwave background (CMB) and baryon acoustic oscillation (BAO) observations can allow for such components before and near recombination. 
   The new dynamical degree of freedom can be early dark energy (EDE) or early modified gravity depending on its coupling to gravity. We study a new model, $\mathcal{G}$EDE, featuring the cubic Galileon operator $X\Box\phi$ and test it against the most recent Planck PR4 CMB and Cepheid calibrated Pantheon+ type Ia Supernovae data. Thanks to the kinetic braiding effects, $\mathcal{G}$EDE gives a better fit to the data, with a higher $H_0$, and is preferred over the canonical EDE with a Bayes factor $\ln B=0.9$, despite introducing one more parameter.  This calls for further explorations of modified gravity near and before last scattering. To facilitate these, we introduce a substantial extension of the cosmological code \texttt{EFTCAMB} that allows to fully evolve  the background and linear dynamics of any covariant theory, oscillatory or not,  belonging to the Horndeski class. 
   
	\end{abstract}
	
	\maketitle
	
	
	\section{Introduction}
   At the core of modern cosmology lies the Hubble constant $H_0$, which sets the size and expansion rate of our current Universe. In the past century tremendous effort has been devoted to measuring this value, both locally or through a cosmological model. The most precise constraint on $H_0$ is derived from the cosmic microwave background (CMB) anisotropy observations assuming the standard cosmological constant ($\Lambda$) cold dark matter (CDM) model. The latest Planck PR4 data report  $H_0=67.81\pm0.38\ \rm{km/s/Mpc^{-1}}$ \cite{Tristram:2023haj}. On the other hand, local observations can also measure $H_0$, by e.g. calibrating the distance-ladder, with the most precise one being the Cepheid calibrated type Ia Supernovae from the SH0ES collaboration, giving $H_0=73.04\pm1.04\ \rm{km/s/Mpc^{-1}}$~\cite{Riess:2021jrx}. 

    The now $5\sigma$ discrepancy, usually referred to as the {\it Hubble tension}, between the local and early Universe determination of $H_0$, has triggered extensive discussions about the possible underlying new physics, see e.g.~\cite{DiValentino:2021izs,Abdalla:2022yfr,Kamionkowski:2022pkx} for recent reviews. Among the plethora of possibilities explored, one of the most promising and extensively studied is the proposal of an additional dark energy component around matter-radiation equality, often dubbed early dark energy (EDE)  ~\cite{Poulin:2018cxd,Agrawal:2019lmo,Alexander:2019rsc,Lin:2019qug,Smith:2019ihp,Niedermann:2019olb,Sakstein:2019fmf,Ye:2020btb,Braglia:2020bym,Karwal:2021vpk,Niedermann:2021vgd,McDonough:2021pdg,McDonough:2022pku,Brissenden:2023yko,Cicoli:2023qri,Wang:2024jug}; see also \cite{Poulin:2023lkg} for a recent review. Stringent constraints on EDE can be derived from large scale structure (LSS) observations~\cite{Hill:2020osr,Ivanov:2020ril,DAmico:2020ods,Ye:2020oix,Pogosian:2020ded,Jedamzik:2020zmd,McDonough:2023qcu,Gsponer:2023wpm}, CMB  data from Planck PR4 and ground based observations \cite{Jiang:2021bab,Poulin:2021bjr,LaPosta:2021pgm,Smith:2022hwi,Cruz:2023cxy,Smith:2023oop,Efstathiou:2023fbn,Ye:2023zel,Khalife:2023qbu}. Despite these constraints, there remains room for  a new dynamical degree of freedom (DoF) at high redshifts, and this is commonly referred to as EDE or early modified gravity (EMG), depending on how it is coupled to gravity. Modifications of the effective Newtonian constant $G_{\rm{eff}}$ have been considered as EMG candidates,  both with dynamical \cite{Zumalacarregui:2020cjh,Braglia:2020auw,FrancoAbellan:2023gec} and parametrical \cite{Lin:2018nxe,Braglia:2020iik,Wen:2023wes} approaches.

    The Generalized Galileons theory~\cite{Deffayet:2011gz} offers a natural,  unifiying framework to explore EDE/EMG models. The corresponding  Lagrangian describes  theories with second order equations of motion (EoM), in a four-dimensional spacetime, for the usual massless graviton plus one additonal scalar DoF.  As shown in~\cite{Kobayashi:2011nu}, this action is equivalent to that of  Horndeski gravity~\cite{Horndeski:1974wa}. As a first step into our exploration of EDE/EMG, we will restrict to the subclass of covariant Galileons,  identified in~\cite{Deffayet:2009wt} as a first generalization of the work of~\cite{Nicolis:2008in} to an expanding background. In particular, we consider the model defined by the following Lagrangian
    \begin{equation}\label{eq:lagrangian}
		L =\frac{M_p^2}{2}R +X-V(\phi)-\xi X\Box\phi.
	\end{equation}
    where $V(\phi)$ is the scalar field potential, $X\equiv-\frac{1}{2}(\partial\phi)^2$ and $\Box\equiv g^{\mu\nu}\nabla_{\mu}\nabla_{\nu}$. The non-canonical dynamics is only characterized by the free parameter $\xi$ multiplying the Galileon operator, $X\Box\phi$. Lagrangian \eqref{eq:lagrangian} corresponds to luminal gravitational waves,  $c^2_T=1$, and a minimally coupled scalar sector, with the only non-trivial modified gravity (MG) effect coming from the kinetic braiding introduced by the last term. Kinetic braiding theories are the majority of Horndeski gravity models that survive the no-ghost and gradient stability conditions as well as positivity bounds~\cite{deBoe:2024gpf}; the latter  encode the fundamental requirement of causality, unitarity and Lorentz invariance of the UV complete theory, however, it is important to note that at the moment they heavily rely on an extrapolation of results from the Minkowski to the cosmological background~\cite{deBoe:2024gpf}.

    The shift symmetric case of~\eqref{eq:lagrangian} has been ruled out as a dark energy candidate by a combination of cosmological observations~\cite{Renk:2017rzu, Peirone:2017vcq}. However, this conclusion assumes shift-symmetry and that the Galileon field sits on its tracking solution, which does not apply to \eqref{eq:lagrangian} with a scalar potential. In the non-shift-symmetric case such as \eqref{eq:lagrangian}, the Galileon operator still offers important MG effect  \eqref{eq:lagrangian} through the kinetic braiding without suffering from stringent constraints. Furthermore, as we will show, with a potential $V(\phi)$ to drive the field evolution, and the expansion of the Universe dominated by radiation and matter, Galileons can be revived as an EDE/EMG theory ($\mathcal{G}$EDE) which is favored at $\sim2\sigma$ over the EDE model based on a canonical scalar field (i.e. setting $\xi=0$ in Eq.\eqref{eq:lagrangian}), hereafter cEDE.

    The paper is structured as follows. In Section~\ref{sec:theory} we present the background and linear perturbation dynamics of $\mathcal{G}$EDE. The numerical setup and the data used are outlined in Section~\ref{sec:data} and the cosmological constraints are presented in Section~\ref{sec:result}. Finally we conclude in Section~\ref{sec:conclusion}.

    \begin{figure}
        \centering
        \includegraphics[width=\linewidth]{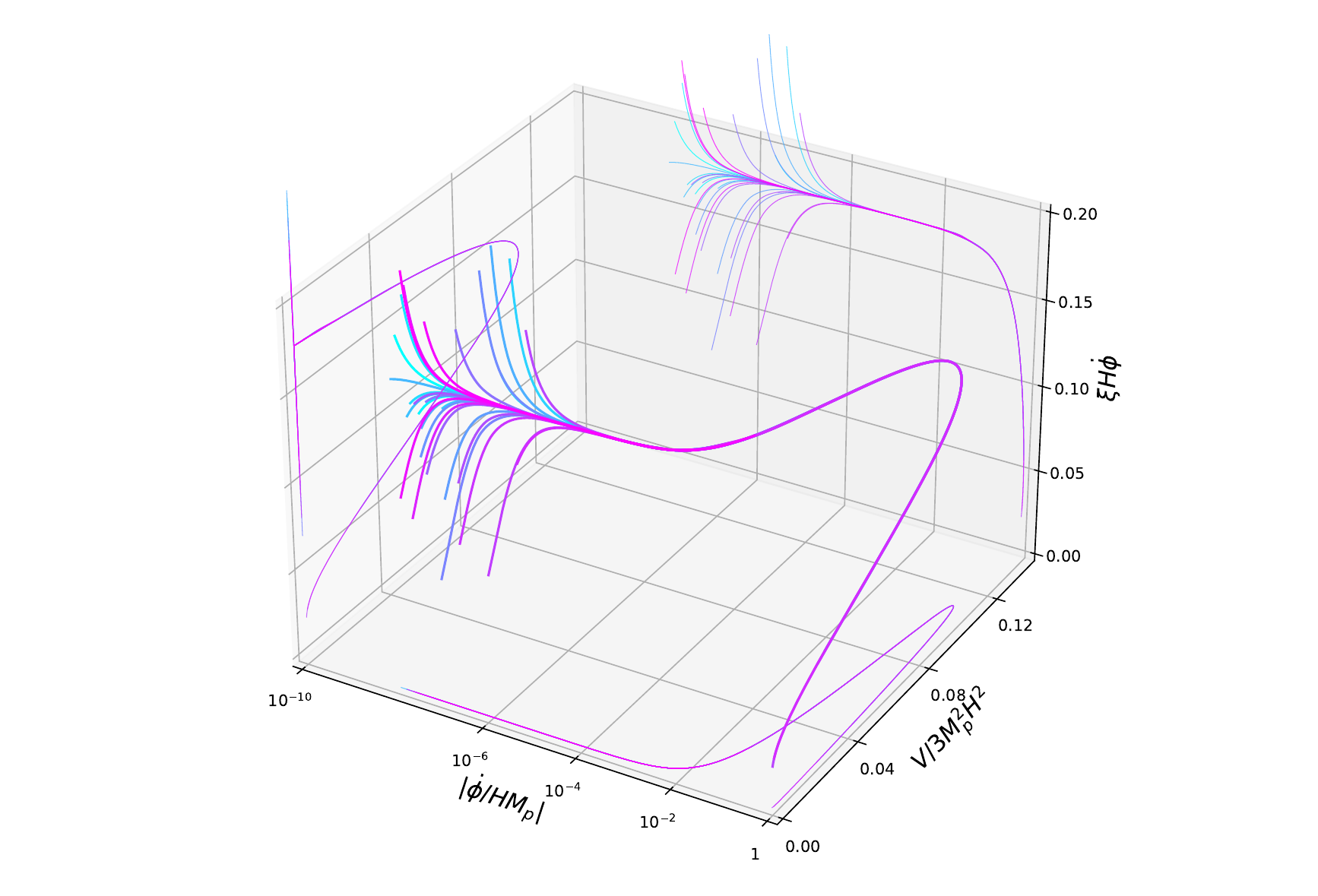}
        \caption{Phase space trajectories of the scalar field in the 3D space spanned by the dimensionless field velocity $\dot{\phi}/HM_p$, the potential energy fraction $V/3M_p^2H^2$ and the Galileon parameter $\xi H\dot{\phi}$. Curves with different colors correspond to different initial conditions, which all converge into the tracking solution. The plot is generated in a $\mathcal{G}$EDE model with a quartic potential and $\xi V_{\phi}=-1.3$.}
        \label{fig:phaseplot}
    \end{figure}
	
	\section{The model}\label{sec:theory}
	The covariant EoM for the scalar field in the $\mathcal{G}$EDE model is
	\begin{equation}\label{eq:eom}
		-\Box\phi + \xi\left[(\Box\phi)^2 - \nabla_{\mu}\nabla_{\nu}\phi\nabla^{\mu}\nabla^{\nu}\phi-R^{\mu\nu}\nabla_{\mu}\phi\nabla_{\nu}\phi\right]+V_{\phi}=0\,.
	\end{equation}
    On a FRW background $ds^2 = -dt^2+a^2(t)|d\bm{x}|^2$, the Einstein equation gives the following Friedmann equations
	\begin{eqnarray}\label{eq:friedmann}
		3 M_p^2 H^2 &=& \rho_m + \frac{1}{2} \dot{\phi}_0^2 + V(\phi_0) + 3 \xi H \dot{\phi}_0^3, \nonumber \\
		2M_p^2 \dot{H} &=& -(\rho_m + p_m + \dot{\phi}_0^2 + 3 \xi H \dot{\phi}_0^3 - \xi \dot{\phi}_0^2 \ddot{\phi}_0 ),
	\end{eqnarray}
	and the scalar field  EoM reduces to 
	\begin{equation}\label{eq:scf_bkeom}
		\ddot{\phi}_0+3H\dot{\phi}_0\left[1+\xi\left(2\ddot{\phi}_0+\left(3+\frac{\dot{H}}{H^2}\right)H\dot{\phi}_0\right)\right]+V_{\phi}=0
	\end{equation}
	where overdot denotes derivation with respect to cosmic time, $t$, and the subscript $0$ denotes background quantities. Similar to cEDE, in $\mathcal{G}$EDE initially the field is frozen at $\phi=\phi_i$ on the potential well by Hubble friction when $m^2_{\phi,i}\sim V_{\phi\phi}(\phi_{i})\ll H^2$. At some point $H$ drops below $m_{\phi,i}$ so the field thaws and rolls down the potential and finally oscillates around the potential minima, rapidly redshifting away its energy density. During the entire process the Gallileon operator, controled by $\xi$, modulates both the background and perturbation evolution though its MG effect. Fig.\ref{fig:phaseplot} depicts the frozen and thawing stage of the dynamical picture.

    The time direction flows from left to right for the trajectories plotted in Fig.\ref{fig:phaseplot}. The initial frozen stage is an attractor and characterized by $\dot{\phi}/HM_p\ll1$, where the field displacement in one Hubble time is very small. Thus we can set $V_\phi\sim V_{\phi,i}\sim const.$ in Eq.\eqref{eq:scf_bkeom} to obtain the approximate tracking solution 
\begin{equation}\label{eq:scf_bktracking}
		H\dot{\phi}_0\simeq\frac{-1+\sqrt{1-\frac{12}{3-\dot{H}/H^2}(\xi V_{\phi,i})}}{6\xi}\simeq \rm{const.}
	\end{equation}
    where we have chosen the branch that reduces to a canonical scalar field $H\dot{\phi}_0\simeq-V_{\phi,i}/(3-\dot{H}/H^2)$ when $\xi\to0$. For compact notation we will drop the subscript $i$ and simply write $\xi V_\phi$ hereafter. This tracking behavior is also evident in Fig.\ref{fig:phaseplot} where $\xi H\dot{\phi}$ remains approximately constant up to the energy peak of the scalar field. Specially, in radiation dominance one has $\dot{H}/H^2=-2$, one will need $\xi V_{\phi}<5/12$ for the square root in Eq.\eqref{eq:scf_bktracking} to be well defined, i.e. the existence of a stable tracking background.
 
    At the linear perturbation level, the dynamics of $\mathcal{G}$EDE can be characterized by the effective functions~\cite{Bellini:2014fua}
	\begin{equation}\label{eq:alphaKB}
		\alpha_K = \frac{ \dot{\phi}_0^2 + 6 \xi H \dot{\phi}_0^3 }{ H^2 M_p^2 }\equiv 6 ( f_X + \alpha_B ) ,\qquad \alpha_B = \frac{\xi \dot{\phi}_0^3 }{H M_p^2}= (6 \xi H \dot{\phi}_0 ) f_X.
	\end{equation}
	where we have defined $f_X \equiv X_0 / (3 M_p^2 H^2)$ as the energy fraction of the canonical kinetic term $X_0 = \dot{\phi}_0^2 / 2$. Compared with cEDE, kinetic braiding is the only MG effect in $\mathcal{G}$EDE, while running of effective Planck mass ($\alpha_M$) and modification of tensor speed ($\alpha_T$) are absent in both models. Using the tracking solution~\eqref{eq:scf_bktracking} and Friedmann equation~\eqref{eq:friedmann}, one can derive the approximate sound speed
	 \begin{equation}\label{eq:cs2early}
	 	c_s^2\simeq\frac{1 + \left( 4 - 2 \dot{H} / H^2 \right) \xi H \dot{\phi}_0 - 3 ( \xi H \dot{\phi}_0 )^2 f_X }{1 + 6 \xi H \dot{\phi}_0 + 9 ( \xi H \dot{\phi}_0 )^2 f_X }
	 \end{equation}
  In the initial tracking stage, gradient stability ($c_s^2>0$) requires $\xi V_\phi < 5/8$ or $\xi V_\phi > 5/6$. Ghost stability ($\alpha_K+3\alpha_B^2/2>0$) yields $\xi V_\phi<5/6$. Combining these conditions with the existence of the background tracking solution \eqref{eq:scf_bktracking} we arrive at the necessary stability condition
	\begin{equation}\label{eq:stability}
		\xi V_{\phi} < 	\frac{5}{12}.
	\end{equation} 
	
	The modified gravity effect is evident only when the scalar field thaws and contributes non-trivially to the total energy budget with $\alpha_{K,B} \propto f_X$ in Eq.~\eqref{eq:alphaKB}. In particular, $\alpha_B=6\xi H\dot{\phi}_0f_X\sim - \xi V_{\phi}f_X$ has the opposite sign as $\xi V_{\phi}$. At the background level, braiding enters the Hubble equation \eqref{eq:friedmann} as
	\begin{equation}
		3M_p^2(1 - \alpha_B)H^2=\rho_m + \frac{1}{2}\dot{\phi}_0^2+V(\phi_0),
	\end{equation}
	thus having $\xi V_{\phi}<0$ (so $\alpha_B>0$) in $\mathcal{G}$EDE can further increase the Hubble parameter compared with a cEDE model with the same $\phi_0(t),\dot{\phi}_0(t)$, meaning smaller sound horizon and enhanced ability to mitigate the Hubble tension. Within the scalar field sound horizon, perturbations are generally suppressed due to scalar field pressure support. The model under consideration introduces also a fifth force on small scales; we can estimate its effect on perturbations via the phenomenological MG functions $\mu$, $\Sigma$ and $\gamma$ which for our model read~\cite{Kase:2018iwp}
	\begin{equation}\label{eq:musigma}
		\mu=\Sigma=1+\frac{\alpha_B^2}{2c_s^2(\alpha_K+3/2\alpha_B^2)},\qquad\gamma=1.
	\end{equation}
	Using Eq.\eqref{eq:alphaKB} and $f_X\ll1$ one has
	\begin{equation}\label{eq:mu-1}
		\mu-1\simeq\frac{12(\xi H\dot{\phi}_0)^2}{1 + \left( 4 - 2 \dot{H} / H^2 \right) \xi H \dot{\phi}_0}f_X\sim\frac{0.48(\xi V_\phi)^2}{1-1.6(\xi V_\phi)}f_X.
	\end{equation}
	In the last approximate equality we have assumed the tracking solution \eqref{eq:scf_bktracking} and radiation dominance. Thus on small scales, the MG effect always enhances gravity and perturbation growth. For superhorizon modes, $\delta\phi$ is not excited so the effect on perturbation evolution outside of horizon is negligible. In general on small scales the MG term enhances perturbation growth, competing with the additional pressure support from the scalar fluid within its sound horizon, which suppresses clustering.
	
	\section{Datasets and Methodology} \label{sec:data}
	In order to perform a fit to the data, we choose an explicit $\mathcal{G}$EDE model by setting the potential to the following quartic ansatz
	\begin{equation}
		V(\phi) = V_0\phi^4+V_\Lambda\,
	\end{equation}
 where $V_\Lambda$ stands for the cosmological constant supporting the late-time accelerated expansion. 
	This potential depends on only one parameter and is thus simpler than the one in the original axion EDE~\cite{Poulin:2018cxd} model, while still being able to considerably alleviate the Hubble tension~\cite{Agrawal:2019lmo, Ye:2020btb}. In general $V_\Lambda$ could also be dynamical to represent the dark energy. The scalar field model is specified by two model parameters $\{\xi, V_0\}$ plus the initial field value $\phi_{ini}$. In practice we use the more intuitive parameters: $f_{\rm{ede}}$ - the peak energy density of the scalar field and $z_c$ - the redshift at which the peak occurs in place of $\{V_0, \phi_i\}$ and use a shooting method to map them back to $\{V_0, \phi_i\}$. Energy density of the scalar field can be ambiguous once MG is included, we thus define
    \begin{equation}
        f_{\rm{EDE}}(z)\equiv 1-\frac{\rho_{\rm{m},tot}(z)}{3M_p^2H^2(z)}
    \end{equation} where $\rho_{\rm{m},tot}$ is the total energy density of all species except for the scalar field. One can also define the energy fraction of the canonical part by $f^c_{\rm{EDE}}\equiv(\dot{\phi}^2/2+V(\phi))/3M_p^2H^2$. For cEDE one has $f^c_{\rm{EDE}}= f_{\rm{EDE}}$ while for $\mathcal{G}$EDE $f^c_{\rm{EDE}}\ne f_{\rm{EDE}}$.
    In light of the discussion in Section.\ref{sec:theory}, instead of $\xi$ we use the dimensionless parameter $\xi V_\phi$ which directly controls the dynamics and stability of the theory. With this setup, we have the standard six cosmological parameters $\{ \omega_b, \omega_c, H_0, \ln 10^{10}A_s, n_s, \tau_{reion} \}$ and three additional model parameters $\{ f_{\rm{ede}}, z_c, \xi V_{\phi} \}$ for our cosmological model. 
 
    To compute the cosmology and the corresponding prediction for the observables of interest, we use a new version of~\texttt{EFTCAMB} \cite{Hu:2013twa,Raveri:2014cka}, based on the public Einstein-Boltzmann code \texttt{CAMB} \cite{Lewis:1999bs,camb}. This version features the implementation of a covariant Horndeski module which evolves \textit{any} covariant Horndeski theory, both at the background and linear level. Provided with an arbitrary Horndeski Lagrangian, the code integrates the scalar field perturbations in terms of $\delta\phi$ instead of $\pi=\delta\phi/\dot{\phi}$, thus avoids the divergence in $\pi$ when $\dot{\phi}$ crosses zero that plagues all previous Boltzmann codes that evolve scalar-tensor theories. The new \texttt{EFTCAMB} is interfaced with \texttt{Cobaya} \cite{Torrado:2020dgo,2019ascl.soft10019T}, which we use to perform Monte Carlo Markov Chain (MCMC) analysis to derive posterior constraints on the model parameters. We use the Gelman-Rubi~\cite{Gelman:1992zz} diagnostic $R-1<0.02$ as our convergence criterium. We also use the nested sampler \texttt{PolyChordLite} \cite{Handley:2015fda,Handley:2015vkr}, interfaced with \texttt{Cobaya}, to compute the Bayesian evidence of the models.
	
	We use the following dataset:
	\begin{itemize}
		\item \textbf{P20}: \texttt{hillipop}  TTTEEE and \texttt{lollipop} EE likelihoods of the Planck2020 CMB temperature and polarization data \cite{Tristram:2023haj} as well as the Planck2018 low-l TT data \cite{Planck:2018vyg}. Planck PR4 CMB lensing \cite{Planck:2019nip}.
		\item \textbf{BAO}: The low redshift BAO from MGS \cite{Ross:2014qpa} and 6dF \cite{Beutler:2011hx} and high redshift BAO data from SDSS DR12 \cite{BOSS:2016wmc}.
		\item \textbf{cSN}: The Pantheon+ Type Ia supernova light curve sample \cite{Brout:2022vxf} sample with SH0ES Cepheid host distance calibration \cite{Riess:2021jrx}.
		\item \textbf{LSS}: Weak lensing shear and galaxy clustering (3x2pt) LSS measurement from the Dark Energy Survey (DES) year one data \cite{DES:2017myr}.
		\item \textbf{Baseline}: P20+BAO+cSN
	\end{itemize}
Due to the significantly increased computational cost, when running nested sampling we replace the high l TTTEEE CMB likelihood with \texttt{hillipop-lite} and the DES Y1 data with a Gaussian prior on $S_8=0.790^{+0.018}_{-0.014}$, from the most recent KiDS and DES combined result \cite{Kilo-DegreeSurvey:2023gfr}.
	
	\section{Results} \label{sec:result}
	\begin{figure}
		\centering
		\includegraphics[width=\linewidth]{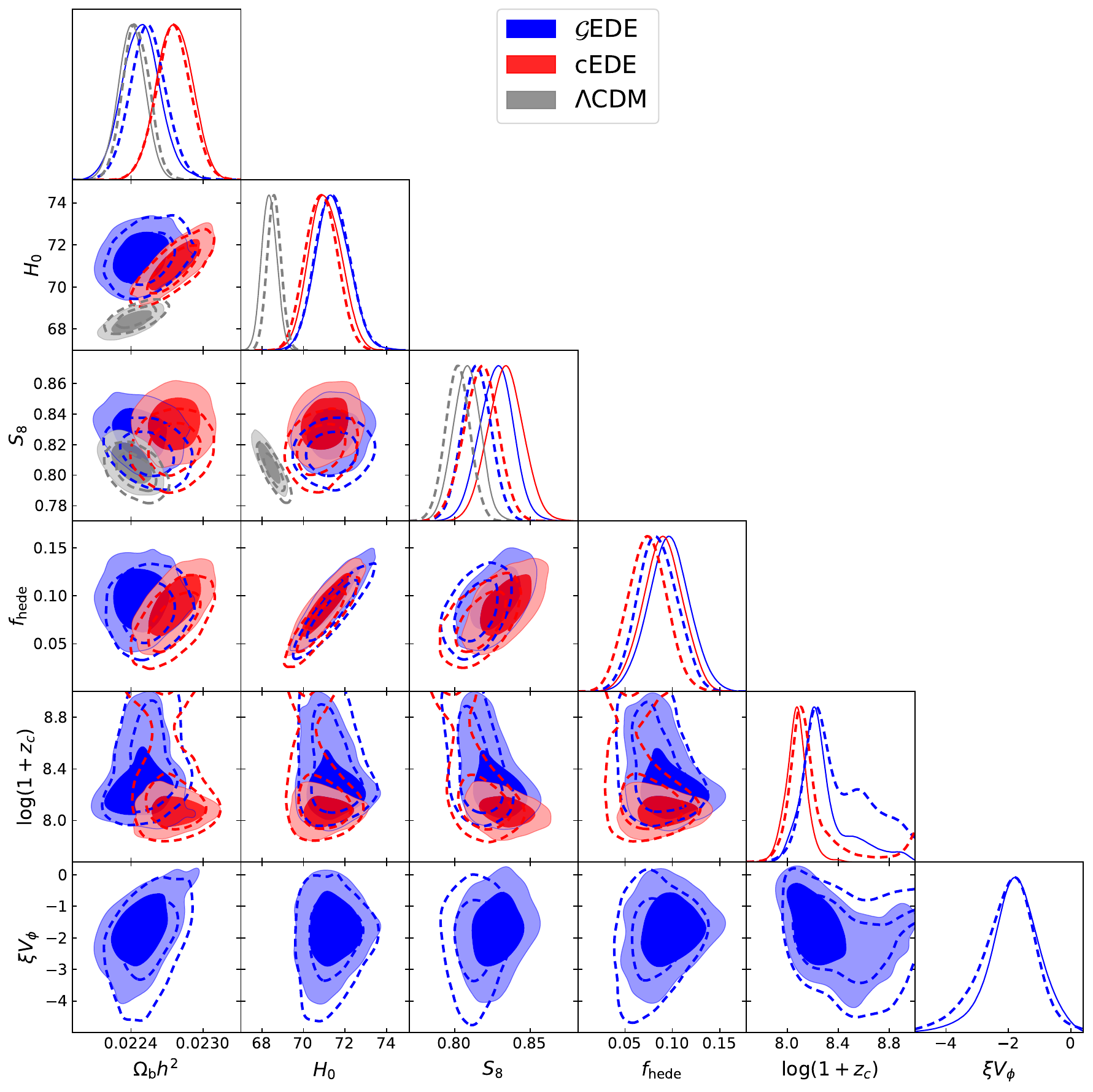}
		\caption{68\% and 95\%   marginalized posterior constraints for selected cosmological and model parameters for $\Lambda$CDM, cEDE, $\mathcal{G}$EDE in the baseline (filled contours) and baseline+LSS (dashed contours) datasets.}
		\label{fig:main_contour}
	\end{figure}
    \begin{table}
        \centering
        \begin{tabular}{|c|c|c|c|c|c|c|}
            \hline 
            Model& \multicolumn{2}{|c|}{$\Lambda$CDM} & \multicolumn{2}{|c|}{cEDE}& \multicolumn{2}{|c|}{$\mathcal{G}$EDE} \\
            \hline
            Dataset& baseline&baseline+LSS& baseline&baseline+LSS& baseline&baseline+LSS\\
            \hline
            $100\omega_b$     &$2.24\pm 0.01  $&$2.24\pm 0.01  $&$2.28\pm 0.02  $&$2.28\pm 0.01        $&$2.25\pm 0.02        $&$2.25\pm 0.02  $ \\
            $100\omega_c$     &$11.74\pm 0.08 $&$11.69\pm 0.07 $&$12.9\pm 0.3   $&$12.6\pm 0.3         $&$12.9\pm 0.3         $&$12.7\pm 0.3$ \\
            $H_0$             &$68.4\pm 0.4   $&$68.6\pm 0.3   $&$71.1\pm 0.8   $&$70.9\pm 0.8         $&$71.4\pm 0.8         $&$71.4\pm 0.8   $ \\
            $n_s$             &$0.972\pm 0.003$&$0.973\pm 0.003$&$0.986\pm 0.005$&$0.984\pm 0.005      $&$0.985\pm 0.005      $&$0.985\pm 0.005$ \\
            $\ln(10^{10}A_s)$ &$3.05\pm 0.01  $&$3.05\pm 0.01  $&$3.07\pm 0.01  $&$3.06\pm 0.01        $&$3.07\pm 0.01        $&$3.06\pm 0.01            $ \\
            $\tau_{\rm{reio}}$&$0.062\pm 0.006$&$0.062\pm 0.006$&$0.062\pm 0.006$&$0.062\pm 0.006      $&$0.061\pm 0.006      $&$0.062\pm 0.006$ \\
            $f_{\rm{hede}}$   &N.A.            & N.A.           &$0.07\pm 0.02  $&$0.07\pm 0.02        $&$0.10\pm 0.02        $&$0.08\pm 0.02  $ \\
            $\ln(1+z_c)$      &N.A.            & N.A.           &$8.07\pm 0.09  $&$8.21^{+0.03}_{-0.22}$&$8.29^{+0.07}_{-0.23}$&$8.4^{+0.2}_{-0.3}      $ \\
            $\xi V_\phi$      &N.A.            & N.A.           & N.A.           & N.A.                 &$-1.8^{+0.8}_{-0.7}  $&$-2.0^{+1.0}_{-0.7}$ \\
            \hline
            $\Omega_m$        &$0.301\pm 0.005$&$0.298\pm 0.004$&$0.302\pm 0.005$&$0.297\pm 0.004      $&$0.299\pm 0.005      $&$0.294\pm 0.004$ \\
            $S_8$             &$0.808\pm 0.009$&$0.802\pm 0.008$&$0.83\pm 0.01  $&$0.82\pm 0.01        $&$0.83\pm 0.01        $&$0.815\pm 0.010$ \\
            \hline
        \end{tabular}
        \caption{68\% posterior constraints on all cosmological parameters.}
        \label{tab:par}
    \end{table}
    \begin{table}
        \centering
        \begin{tabular}{|c|c|c|c|c|c|c|}
            \hline 
            Model& \multicolumn{2}{|c|}{$\Lambda$CDM} & \multicolumn{2}{|c|}{cEDE}& \multicolumn{2}{|c|}{$\mathcal{G}$EDE} \\
            \hline
            Dataset& baseline&baseline+LSS& baseline&baseline+LSS& baseline&baseline+LSS\\
            \hline
             $\chi^2_{\rm{CMB}}$ &30561.4 &30559.8 &30561.9 &30561.2 &30560.3 &30559.4 \\
             $\chi^2_{\rm{CMB\ Lensing}}$ &8.6 &10.0 &9.1 &8.9 &9.4 &10.5 \\
             $\chi^2_{\rm{cSN}}$ &1484.2 &1485.8 &1463.2 &1464.7 &1459.1 &1459.5 \\
             $\chi^2_{\rm{BAO}}$ &5.9 &5.5 &5.4 &5.6 &5.5 &6.7 \\
             $\chi^2_{\rm{LSS}}$ &N.A. &506.1 &N.A. &508.3 &N.A. &507.5 \\
             \hline
             $\chi^2_{\rm{tot}}$ &305047.4 &32599.5 &32039.6 &32548.6 &32034.3 &32543.6 \\
             $\ln Z$ & N.A. &-3592.7& N.A. &-3586.8 & N.A. &-3585.9 \\
            \hline
        \end{tabular}
        \caption{Bestfit $\chi^2$ per dataset, and total, for the three different models. The last line reports the Bayesian evidence of the corresponding model obtained from nested sampling.}
        \label{tab:bestfit_chi2}
    \end{table}
	We show the marginalized posterior distributions of relevant model parameters in Fig.~\ref{fig:main_contour}, and report the posterior constraints for all parameters in  Table.~\ref{tab:par}. Additionally, we report the full posterior results in Appendix-\ref{apdx:fullmcmc}. Finally, in Table.~\ref{tab:bestfit_chi2} we report the  $\chi^2$ for the bestfit model, total as well as per experiment, and the bayesian evidence. There is a significant $\Delta\chi^2 \ (\gtrsim20)$ in the cSN dataset between both EDE models and $\Lambda$CDM. This is because $\Lambda$CDM is in strong Hubble tension with the cSN data while the EDE models greatly alleviate that. 
 
Looking at Fig.~\ref{fig:main_contour}, we see that both cEDE and $\mathcal{G}$EDE show shifts in the  values of $n_s,\omega_m$ compared with $\Lambda$CDM as observed and explained in~\cite{Ye:2020oix,Ye:2021nej}, but $\mathcal{G}$EDE predicts an $\omega_b$ comparable to $\Lambda$CDM, with the central value $\sim2\sigma$ lower than cEDE. The Hubble tension is mitigated with the inclusion of the dark energy field, while presence of the MG term $X\Box\phi$ further improves the fit to the cSN data with $\Delta\chi^2_{\rm{SN}}\simeq-5$ as well as the CMB data with $\Delta\chi^2_{\rm{CMB}}\simeq-2$, bringing the latter to the same goodness-of-fit as $\Lambda$CDM in terms of CMB. Due to the overall improved fit, we observe a preference for the MG term, i.e. $\xi V_\phi \ne 0$, at $2\sigma$ (95\% C.L.). In terms of Bayesian evidence, both dark energy models are strongly favored over $\Lambda$CDM with Bayes factor $\ln B \equiv \ln (Z_1/Z_0) > 6 $ due to the alleviation of the Hubble tension. Moreover, despite one more free parameter, $\mathcal{G}$EDE is still favored over the cEDE with $\ln B=0.9$, corresponding to a $\sim 1\sigma$ preference.

    \begin{figure}
        \centering
        \includegraphics[width=0.48\linewidth]{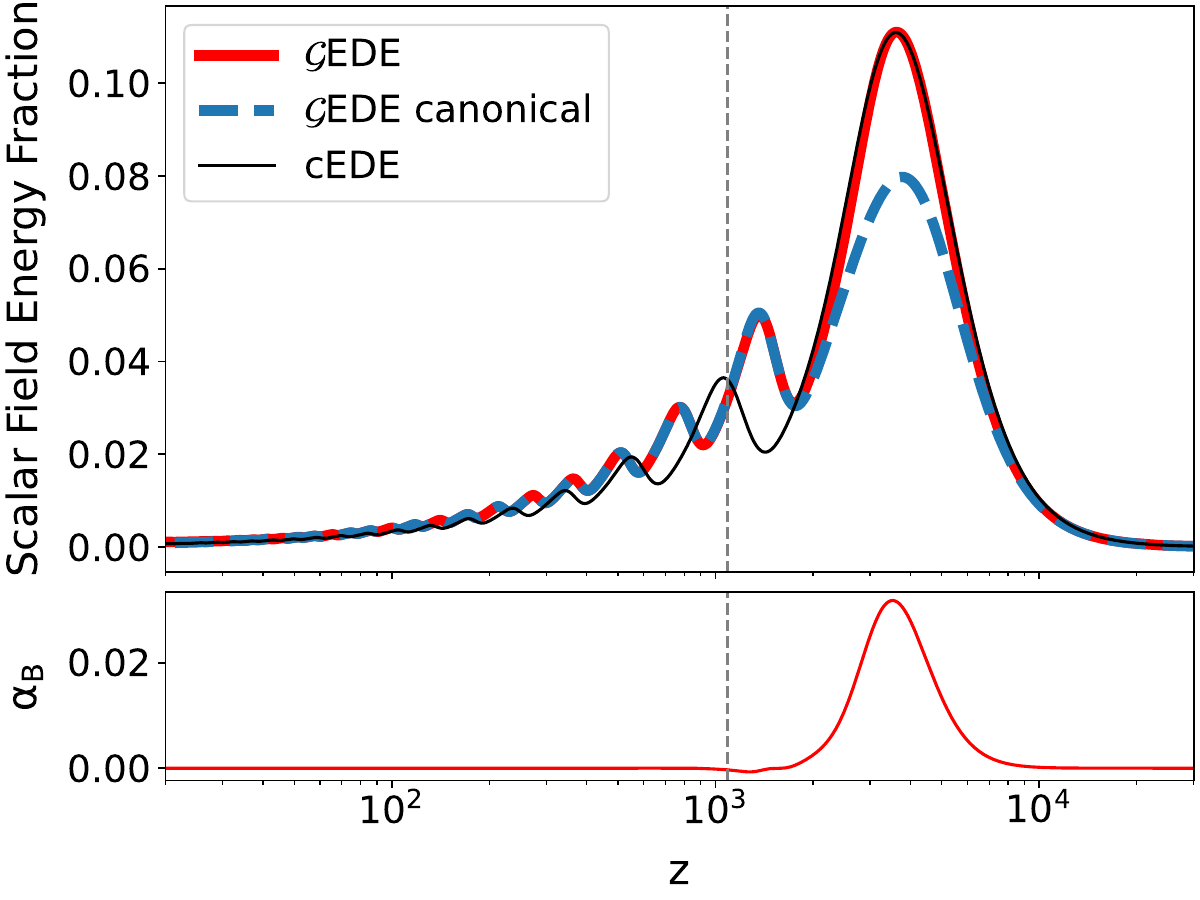}
        \includegraphics[width=0.48\linewidth]{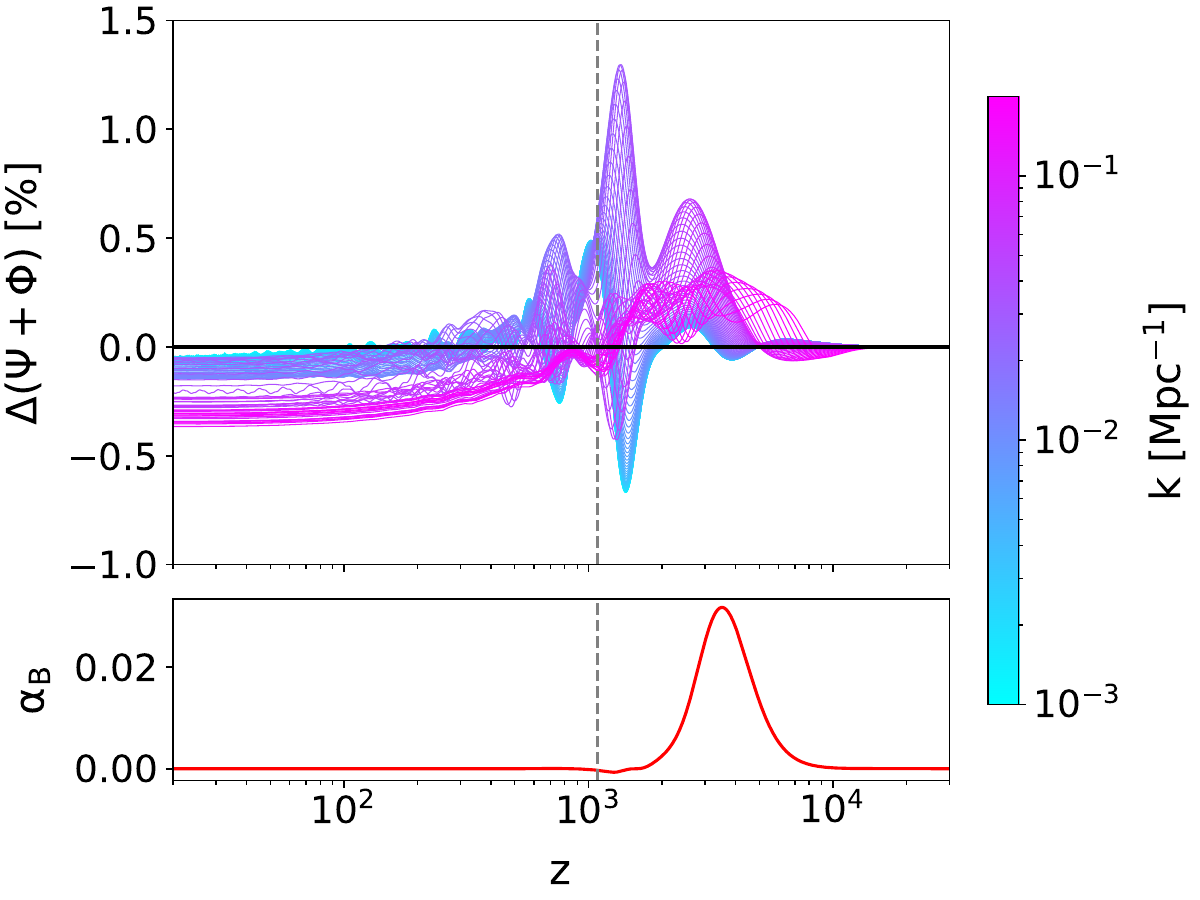}
        \caption{
        Comparison between cEDE and $\mathcal{G}$EDE with the same set of cosmological and model ($\{f_{\rm{EDE}}, z_c\}$) parameters, except for $\xi=0$ in cEDE. The recombination time is marked by a vertical gray dashed line. In the lower panels of both figures we plot the kinetic braiding parameter $\alpha_B$, which characterizes the size of the MG effect in $\mathcal{G}$EDE. \textit{Left panel:} Scalar field energy fraction defined as $f_{\rm{EDE}}\equiv 1-\rho_{\rm{m,tot}}/(3M_p^2H^2)$ where $\rho_{\rm{m,tot}}$ is the total energy density of all species except for the early dark energy field. The blue dashed line ($\mathcal{G}$EDE canonical) plots the canonical part of scalar field energy density faction $f^c_{\rm{EDE}}\equiv(\dot{\phi}^2/2+V(\phi))/3M_p^2H^2$ in $\mathcal{G}$EDE, which is significantly smaller than the full quantity at the peak due to kinetic braiding. \textit{Right panel:} Landscape of fractional difference in the Weyl potential compared with cEDE. The plotted value is $(\Psi+\Phi)_{\rm{MG}}/(\Psi+\Phi)_{\rm{Base}}-1$. }
        \label{fig:bkpt}
    \end{figure}
    
    \begin{figure}
        \centering
        \includegraphics[width=0.9\linewidth]{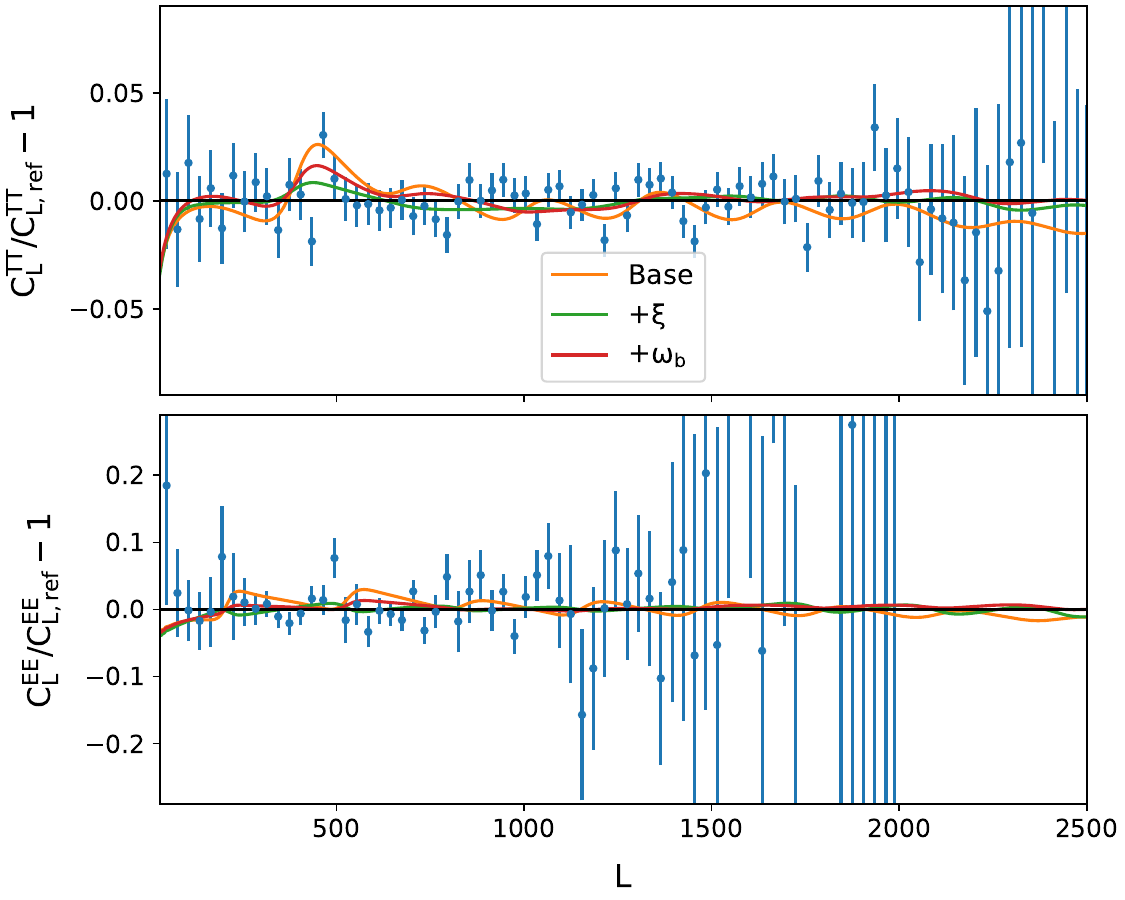}
        \caption{
        Fractional difference in the TT and EE spectra with respect to the Planck18 bestfit $\Lambda$CDM mdoel. Binned data points are from Planck18. ``Base" is the bestfit $\mathcal{G}$EDE model but with $\xi=0$. As explained in the main text, without the MG term it produces less power at high-$\ell$. To compensate, one can either increase baryon density (the ``$+\omega_b$" line) or turn on MG (the ``$+\xi$" line), where data turned out to favor the latter over the former.
        }
        \label{fig:ttee}
    \end{figure}
    
    The $2\sigma$ evidence for $\xi\ne0$ comes both from features at the background and perturbation levels. At the background, $\mathcal{G}$EDE predicts a slightly but consistently larger $H_0$ than cEDE, while fitting better with cSN with $\Delta\chi^2\simeq-5$. This is due to the kinetic braiding effect in the Friedmann equation \eqref{eq:friedmann}, where a $\xi V_\phi<0$, thus a $\alpha_B>0$, enhances the Hubble parameter compared to a canonical field. This is clearly seen in the left panel of Fig.\ref{fig:bkpt} where $f_{\rm{EDE}}$ in $\mathcal{G}$EDE is significantly boosted by $\alpha_B>0$ compared with its canonical part $f^c_{\rm{EDE}}$. Note we have $\alpha_{B,\rm{max}}\simeq0.03$, which enhances the Hubble parameter by 1.5\%. However, because the scalar field only takes up at most $\sim10\%$ of the total energy budget, the small increase in Hubble translates to a significant boost ($\sim 20\%$) in the scalar field peak energy fraction.  At the level of perturbations, $\mathcal{G}$EDE improves the fit to the CMB spectra with $\Delta\chi^2_{\rm{CMB}}\simeq-2$. It has been identified in \cite{Ye:2021nej} that an increase $\omega_b$ is needed to shrink the physical damping scale in all cEDE models to prevent the damping angular scale from changing drastically when $H_0$ is increased, but has the side effect of also impacting the baryon drag which changes the relative height between even and odd acoustic peaks, requiring imperfect compensation from other sources, including tuning $n_s$ \cite{Ye:2021nej} and early integrated Sachs-Wolfe (ISW) effect \cite{Vagnozzi:2021gjh}. In contrast, $\mathcal{G}$EDE completely negates such $\omega_b$ shift in Fig.\ref{fig:main_contour} with $\xi V_\phi\sim-2$, indicating a new degeneracy direction between the MG effect and a larger angular damping  scale. Fig.\ref{fig:ttee} compares the effect of $\omega_b$ and MG in the CMB power spectra. We setup the ``Base" model as the bestfit $\mathcal{G}$EDE but with $\xi=0$, in which the TT spectrum is overdamped at high-$\ell$. Turning on the Galileon operator (``$+\xi$") or increasing baryon density (``$+\omega_b$") can both compensate for the excess damping, but increasing $\omega_b$ suffers from modified baryon drag around $\ell\sim500$. In $\mathcal{G}$EDE, MG adds power to the high-$\ell$ tail by enhancing gravity on small scales, see right panel of Fig.\ref{fig:bkpt}, and thus radiation driving. This effect might be general to all MG with an actractive fifth force, as in Eq.\eqref{eq:mu-1}. Near horizon scale, the MG effect on perturbations in Fig.\ref{fig:bkpt} is difficult to clarify analytically due to the many factors at play. However, the change in baryon drag effect due to a shift in $\omega_b$ is of order $\delta\omega_b/\omega_b|\Psi|\sim0.03|\Psi|$ in the first two peaks, but at such scales radiation driving is already subdominant and the change in early ISW, according to Fig.\ref{fig:bkpt}, is only at sub percent level. Therefore, the impact of MG on intermediate scales is generally smaller than $\omega_b$ thus the better fit.
    
    The $S_8$ problem is not alleviated by $\mathcal{G}$EDE, nor does it get worse. However, the MG model does predict slightly smaller $S_8$ than the canonical model. We attribute this reduction mainly to the reduced $\omega_b$ thus a smaller $\Omega_m$. MG does not help to this end because the fifth force is always attractive at small scales according to Eq.\eqref{eq:musigma}, but it also does not worsen the problem because $\mu-1>0$ is very small, e.g. with $\xi V_\phi=-2$ and $f_X=0.1$ Eq.\eqref{eq:mu-1} gives $\mu-1<5\%$, and only non-vanishing in a narrow redshift window when the scalar field energy fraction is non-negligible. To explicitly help with the $S_8$ tension one might need a repulsive fifth force ($\mu<1$), which, according to \cite{Kase:2018iwp}, requires non-trivial $G_4$ in Horndeski theories. It has been shown that $G_4\supset\phi^2$ might have some effect but does not differ much from the general prediction of EDE, i.e. increased $S_8$ than $\Lambda$CDM \cite{Braglia:2020auw,FrancoAbellan:2023gec}. In Fig.\ref{fig:main_contour}, adding LSS (dashed contours) considerably widens some posterior distributions, especially that of $z_c$. With LSS, there are MCMC points accumulate at the upper prior boundary $\ln(1+z_c)<9$, creating a secondary peak in the distributions. It is a known feature of EDE that the model displays bimoduality in $z_c$ with the global bestfit living in the main peak with lower $z_c$, see e.g. \cite{Karwal:2024qpt}. The high $z_c$ subpeak fits worse to the baseline data but predicts a smaller $S_8$, thus once LSS is included, the competition between data in tension degrades the constraints and drives the contour to extend towards the high $z_c$ region. Appendix-\ref{apdx:hz_results} further explores the high redshift model by excluding the low $z_c$ peak from the prior range.
	
	\section{Conclusion} \label{sec:conclusion}
    We have studied the phenomenology of the Galileon field with a non-canonical operator $X\Box\phi$ as an realization of EMG/EDE, with the resultant model dubbed $\mathcal{G}$EDE. It is found that, depite having one more parameter, $\mathcal{G}$EDE is still favored over canonical EDE with a Bayes factor $\ln B=0.9$ due to better fit to the CMB and SNIa data. We identify the source of improvement as contribution from the MG effect (kinetic braiding) on both background and linear perturbation levels. It's known that noticeable increase in $\omega_c$, $\omega_b$ and $n_s$ is a common feature of EDE models in order to fit CMB and BAO \cite{Ye:2020oix,Ye:2021nej}. The necessity of increasing $\omega_b$ is removed in $\mathcal{G}$EDE due to the small scale attractive fifth force, resulting in improved consistency with small scale CMB observations as well as recent BBN constraints \cite{Mossa:2020gjc}. Aside from $\omega_b$, $n_s$ and $\omega_c$ in $\mathcal{G}$EDE still follow the same increasing trend as in previous EDE/EMG models. The increase in $\omega_c$ is related to the $S_8$ tension which plagues $\mathcal{G}$EDE as well. The increase in $n_s$ connects the Hubble tension with a Harrison-Zeldovich initial spectrum \cite{Ye:2021nej,Ye:2022afu,Jiang:2022uyg,Jiang:2022qlj,Jiang:2023bsz}, which might be of profound implication in the study of the primordial Universe \cite{Kallosh:2022ggf,Ye:2022efx,Fu:2023tfo,Cecchini:2024xoq,Giare:2024akf}.

     It should, however, be stressed that Lagragian \eqref{eq:lagrangian} is very simplified from a theoretic point of view. For example, the theory will additionally include non-minimal coupling to gravity, i.e. $G_4(\phi)$, if regarded as a direct derivation of local modification of gravity \cite{Nicolis:2008in,Chow:2009fm}. Moreover, it will violate the positivity bounds postulated in~\cite{Tokuda:2020mlf, deBoe:2024gpf} if $V_{\phi\phi}<0$, e.g. in the axion-like EDE potential \cite{Poulin:2018cxd} before the field thaws. One simple extension to remedy is adding a positive non-standard kinetic term $X^2$. The possible extensions are left for future study.

    Another attractive possibility is kinetic non-minimal coupling $G_4(X)$. Because EDE is negligible after recombination, such models evade the $c_T=1$ constraint from GW170817 \cite{LIGOScientific:2017vwq,LIGOScientific:2017zic,Coulter:2017wya} while also contributing non-trivially to CMB observables, including B-mode polarization. Furthermore, as recently pointed out by \cite{Ye:2023xyr}, EDE/EMG can also leave unique resonant signatures in the stochastic gravitational wave background if the field is oscillating, which is a ubiquitous feature of many EMG/EDE models. Resonance in the scalar sector \cite{Smith:2023fob} is also worth studying . 
	
	\begin{acknowledgments}
		This work is supported by NWO and the Dutch Ministry of Education, Culture and Science (OCW) (grant VI.Vidi.192.069). Some of the plots are made with the help of \texttt{GetDist} \cite{2019arXiv191013970L}. The authors acknowledge the ALICE and Xmaris clusters for computational support. The cosmology code used to produce the results is a major component of the new \texttt{EFTCAMB} (in preparation) which will be made public later this year. Currently the code can be provided upon reasonable request. 
	\end{acknowledgments}
	
	\appendix
	
	\section{Supplementary MCMC details}\label{apdx:fullmcmc}
    \begin{figure}
        \centering
        \includegraphics[width=\linewidth]{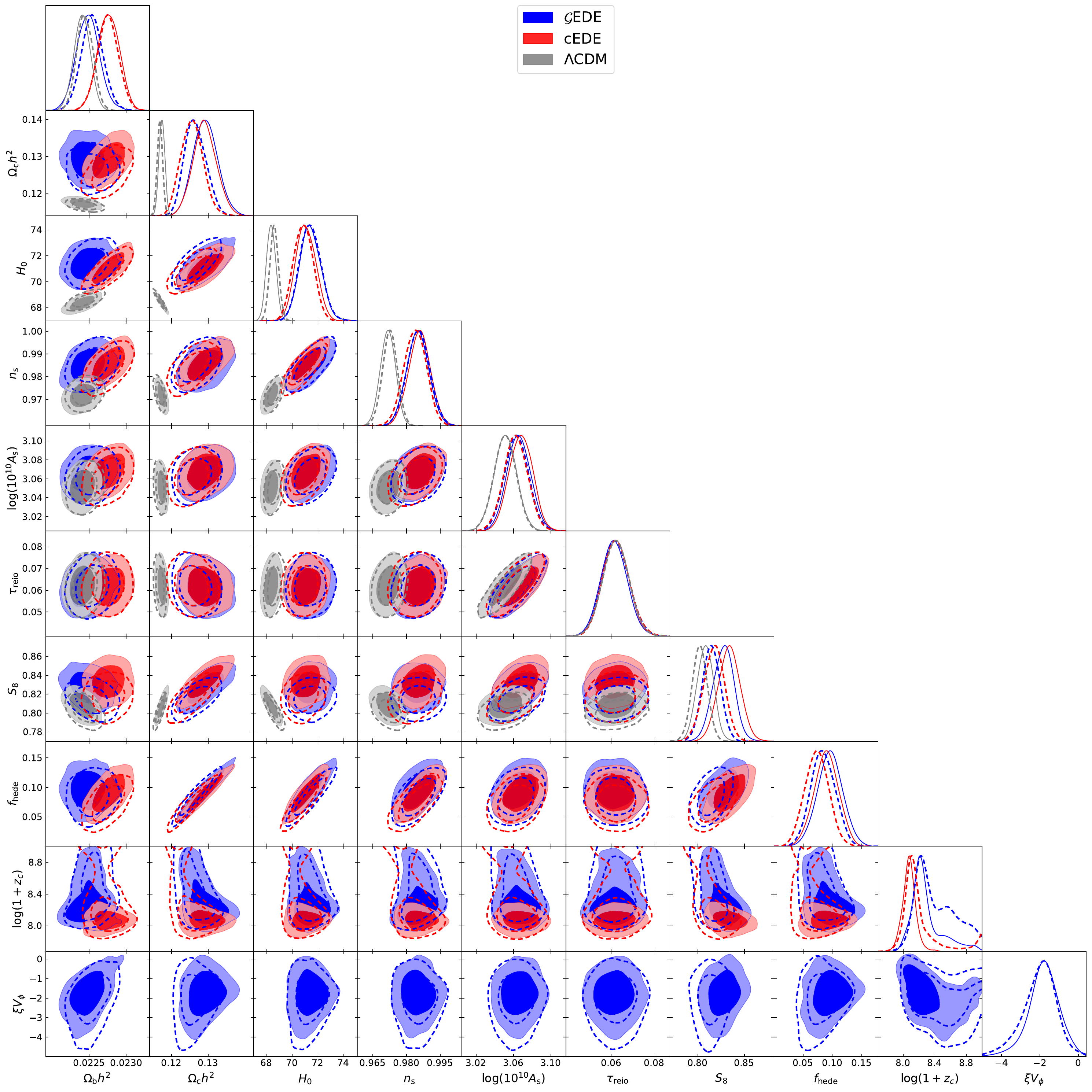}
        \caption{68\% and 95\% marginalized posterior constraints of for all cosmological and model parameters in $\Lambda$CDM (gray), cEDE (red) and $\mathcal{G}$EDE (blue) with the baseline (filled) and baseline+LSS (dashed lines) datasets.}
        \label{fig:full_contour}
    \end{figure}
    \begin{table}
        \centering
        \begin{tabular}{|c|c|c|c|c|c|c|}
            \hline 
            Model& \multicolumn{2}{|c|}{$\Lambda$CDM} & \multicolumn{2}{|c|}{cEDE}& \multicolumn{2}{|c|}{$\mathcal{G}$EDE} \\
            \hline
            Dataset& baseline&baseline+LSS& baseline&baseline+LSS& baseline&baseline+LSS\\
            \hline
            $100\omega_b$        &$2.249$&$2.242$  &$2.276$ &$2.290$ &$2.243$ &$2.261$  \\
            $100\omega_c$        &$11.69$&$11.72$  &$12.79$ &$12.61$ &$13.17$ &$12.97$  \\
            $H_0$                &$68.60$&$68.41$  &$70.99$ &$70.82$ &$71.81$ &$72.22$  \\
            $n_s$                &$0.9750$&$0.9730$&$0.9868$&$0.9845$&$0.9867$&$0.9879$ \\
            $\ln(10^{10}A_s)$    &$3.061$ &$3.047$ &$3.072$ &$3.070$ &$3.070$ &$3.068$  \\
            $\tau_{\rm{reio}}$   &$0.0649$&$0.0619$&$0.0640$&$0.0652$&$0.0627$&$0.0633$ \\
            $f_{\rm{hede}}$      & N.A.   & N.A.   &$0.083$ &$0.075$ &$0.111$ &$0.106$  \\
            $\ln(1+z_c)$         & N.A.   & N.A.   &$8.02$  &$8.11$  &$8.20$  &$8.20$   \\
            $\xi V_\phi$         & N.A.   & N.A.   & N.A.   & N.A.   &$-1.77$ &$-1.61$  \\
            \hline
            $\Omega_m$           &$0.2976$&$0.2998$&$0.3002$&$0.2983$&$0.3003$&$0.2931$ \\
            $S_8$                &$0.810$ &$0.805$ &$0.832$ &$0.824$ &$0.835$ &$0.820$  \\
            \hline
        \end{tabular}
        \caption{Bestfit values of of all cosmological and model parameters in $\Lambda$CDM, cEDE and $\mathcal{G}$EDE with the baseline and baseline+LSS datasets.}
        \label{tab:bestfit_par}
    \end{table}
    \begin{table}
        \centering
        \begin{tabular}{|c|c|c|}
            \hline
                              & MCMC            & NS             \\
            \hline
            $\omega_b$        &$[0.01, 0.03]$ &$[0.02, 0.025]$ \\
            $\omega_c$        &$[0.10, 0.15]$ &$[0.1, 0.15]$   \\
            $H_0$             &$[60, 80]$     &$[60, 80]$      \\
            $n_s$             &$[0.8, 1.2]$   &$[0.9, 1.1]$    \\
            $\ln(10^{10}A_s)$ &$[1.61, 3.91]$ &$[3.0, 3.1]$    \\
            $\tau_{\rm{reio}}$&$[0.01, 0.8]$  &$[0.04, 0.1]$   \\
            $f_{\rm{hede}}$   &$[0, 0.3]$     &$[0, 0.2]$      \\
            $\ln(1+z_c)$      &$[7, 9]$       &$[7, 10]$       \\
            $\xi V_\phi$      &$[-5, 0.4]$    &$[-5, 0.4]$     \\
            \hline
        \end{tabular}
        \caption{Uniform prior used for cosmological and model parameters in the MCMC and nested sampling (NS) methods.}
        \label{tab:prior}
    \end{table}
    Fig.\ref{fig:full_contour} shows the posterior distributions of all comoloigcal and model parameters of $\Lambda$CDM, cEDE and $\mathcal{G}$EDE in the baseline (filled) and baseline+LSS (dashed line) datasets. The corresponding bestfit values, obtained from the minimization module in \texttt{Cobaya}, are summarised in Table.\ref{tab:bestfit_par}. The sampled cosmological parameters are $\{ \omega_b, \omega_c, H_0, \ln 10^{10}A_s, n_s, \tau_{reion} \}$, namely the baryon and cold dark matter density parameter $\omega_{c,b}\equiv\Omega_{c,b}h^2$, the Hubble constant $H_0$, the log amplitude $\ln 10^{10}A_s$ and spectra tilt $n_s$ of the primordial curvature perturbations and the effective reionization optical depth $\tau_{reion}$. The cEDE model adds two more model parameters $\{f_{\rm{ede}}, \ln(1+z_c)\}$ to be sampled, namely the scalar field peak energy fraction $f_{\rm{ede}}$ and logarithm of its redshift postion $\ln(1+z_c)$. 
    Following Planck \cite{Planck:2018vyg} we treat the neutrinos as two massless ulra-relativistic species with $N_{ur}=2.0308$ plus one massive with mass 0.06eV, reproducing $N_{eff}=3.044$ \cite{Bennett:2020zkv,Froustey:2020mcq}. Together with the cosmological and model parameters we also sample all nuisance parameters of the corresponding likelihoods with their recommended priors. We use uninformative priors for cosmological and model parameters as reported in Table.\ref{tab:prior}. The upper bound on $\xi V_{\phi}$ is informed by the theoretical stability requirement \eqref{eq:stability}. We have checked that relaxing this prior will not meaningfully change the posterior distribution because the theories with $\xi V_{\phi}>5/12$ are automatically discarded due to numerical instability, but enforcing the bound in prior makes the MCMC runs much more numerically robust. Due to the significantly increased numerical cost, we shrink some of the priors in nested sampling to speed up convergence, with the exception of $\ln(1+z_c)$ because nested sampling works well with bimodality.

    \section{Constraints on the high-$z_c$ models}\label{apdx:hz_results}
    \begin{figure}[h]
        \centering
        \includegraphics[width=\linewidth]{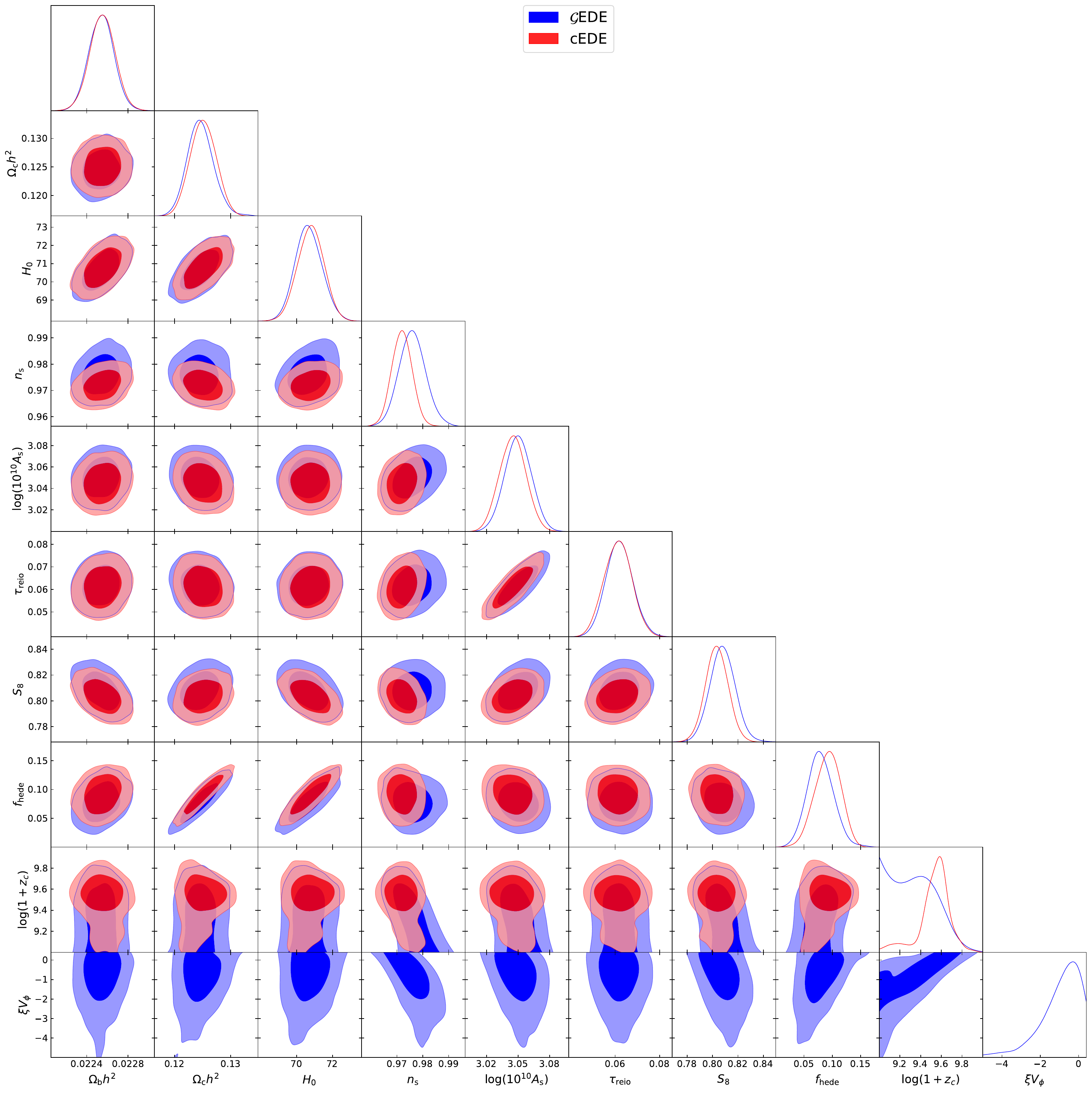}
        \caption{68\% and 95\%   marginalized posterior constraints for all cosmological and model parameters in the baseline cEDE (red) and $\mathcal{G}$EDE (blue) models with prior $9<\ln(1+z_c)<10$.}
        \label{fig:hz_contour}
    \end{figure}
    We explore in this appendix the high redshift secondary peak of cEDE and $\mathcal{G}$EDE. With the same setup as in Appendix-\ref{apdx:fullmcmc} but changing the prior of $\ln(1+z_c)$ to $[9,10]$, we obtain the baseline MCMC posterior constraints as shown in Fig.\ref{fig:hz_contour}. The cEDE model converges to the high $z_c$ minima while turning on Galileon MG in $\mathcal{G}$EDE spoils the convergence. We attribute this to the fact that the new dimension ($\xi V_\phi$) connects the high redshift local minima with the global one around matter-radiation equality by lowering the $\chi^2$ barrier between the two minimas, thus the degeneracy band in the $\xi V_\phi - \ln(1+z_c)$ panel. In fact, in the nested sampling runs, this high-$z_c$ local minima is covered by our prior range, see Table.\ref{tab:prior}, and has been correctly identified, but turned out to be disfavored in both cEDE and $\mathcal{G}$EDE compared with the low redshift minima studied in the main text.
		
	\bibliography{reference}
		
\end{document}